\documentclass[a4paper]{jpconf}
\usepackage{graphicx}

\usepackage{amssymb, amsmath}
\usepackage[final]{epsfig}
\usepackage{graphicx}
\usepackage{graphics}
\usepackage{float}
\usepackage{subfigure}
\usepackage{amsmath,epsfig}
\usepackage[all]{xy}
\begin{document}
\title{Coexistence of  Near-Field and Far-Field Sources: the Angular Resolution Limit }

\author{R\'emy Boyer$^{\natural}$, Mohammed Nabil EL Korso$^{\sharp}$, Alexandre Renaux$^{\natural}$ and Sylvie Marcos$^{\natural}$}

{ \address{$^{\natural}$Laboratoire des Signaux et Systemes (L2S)
\hfill $^{\sharp}$Laboratoire  SATIE, CNRS   \\ Universit\'e
Paris-Sud XI
(UPS), CNRS, SUPELEC, \hfill \'Ecole normale supérieure de Cachan \\
Gif-Sur-Yvette, France \hfill  Cachan, France}
  }

\ead{remy.boyer@lss.supelec.fr}

\begin{abstract}
Passive source localization is a well known inverse problem in
which
  we convert the
observed measurements into information about the direction of
arrivals. In this paper we focus on the optimal resolution of such
problem. More precisely, we propose in this contribution to derive
and analyze the Angular Resolution Limit (ARL) for the scenario of
mixed Near-Field (NF) and Far-Field (FF) Sources. This scenario is
relevant to some realistic situations. We base our analysis on the
Smith's equation which involves the Cram\'er-Rao Bound (CRB). This
equation provides the theoretical ARL which is independent of a
specific estimator. Our methodology is the following: first, we
derive a closed-form expression of the CRB for the considered
problem. Using these expressions, we can rewrite the  Smith's
equation as a 4-th order polynomial by assuming a small separation
of the sources. Finally, we derive in closed-form the analytic ARL
under or not the assumption of  low noise variance. The obtained
expression is compact and can provide useful qualitative
informations on the behavior of the ARL.
\end{abstract}

\section{Introduction}

Very few works are related to the study of the realistic situation
where there exists coexisting far-field (FF) and near-field (NF)
sources \cite{publi1} such as speaker localization using
microphone arrays and guidance (homing) systems. At the contrary,
we can find a plethora of contributions on the  localization of
far-field sources \cite{krim}. More recently, the problem of
localization of  near-field sources has been tackled in reference
\cite{nfl,abed} for instance. In the context of the problem of
source localization, one can see three contributions: (1) propose
new efficient algorithms/estimators \cite{krim}, (2) study the
estimation performance independently of a specific algorithm
thanks to the lower bound on the Mean Square error (MSE)
\cite{stoica_book,abed} and  (3) derive and study the theoretical
resolution, {\em i.e.}, the minimal angular distance to
resolve/discriminate two closely spaced emitted signals in terms
of their direction of arrivals. Our contribution belongs to the
third point. More precisely,  based on the Smith's equation
\cite{smith,1,2} which involves the Cram\'er-Rao Bound (CRB)
\cite{stoica_book}, we derive and analyze the Angular Resolution
Limit (ARL) for  the realistic scenario where we have two sources,
one located in the far-field of the array and another in the
near-field of the array.

\section{Model setup}

We consider some practical applications where the signals $s_1(t)$
and $s_2(t)$ with $t \in [1:T]$ in which $T$ is the number of
snapshots received by an uniform linear array composed by $L$
sensors are the mixture of near-field and far-field sources. More
precisely, the $(TL) \times 1$ observation vector is defined as $
{\bf y} =  {\bf \bar{y}} +  {\bf e} $ where ${\bf e}$ is the
complex centered circular additive white Gaussian noise of
variance $\sigma^2$ and the  $(TL) \times 1$ noise-free signal is
$
  {\bf \bar{y}}  = \begin{bmatrix} {\bf A} {\bf s}(1) \\ \vdots \\ {\bf A} {\bf s}(T) \end{bmatrix}$ with $ {\bf s}(t)  = [s_1(t) \ s_2(t)]^T$ and $
 {\bf A}  = \begin{bmatrix} {\bf a}(\omega_1)  &{\bf b}(\omega_2,\phi)  \end{bmatrix}$ where the signal sources $s_m(t)$ are viewed as deterministic known signals and the steering vectors are defined by $[{\bf a}(\omega_1)]_\ell  =  e^{i \omega_1 \ell}$ and $[{\bf b}(\omega_2,\phi) ]_\ell= e^{i (\omega_2 \ell + \phi \ell^2)}$ with $\ell \in [0:L-1]$. In addition,  we assume $\omega_1\neq \omega_2$. We define the determinist non-zero separation by $\delta=\omega_2-\omega_1$. We focus our attention on the electric parameters but these parameters can be linked to the physical parameters, namely the DOA $\theta_m$ and the range $r$, according to $\omega_m = -2 \pi d/\lambda \sin(\theta_m)$ and $\phi= \pi d^2 /(\lambda r) \cos(\theta_2)^2$ where $d$ is the distance inter-sensor and $\lambda$ is the wavelength.

\section{Analytic ARL in the Smith's sense}

\subsection{Analytical expression of the Cramer-Rao Bound}

The Cramer-Rao Bound (CRB) verifies the covariance inequality principle \cite{stoica_book}. This bound is largely used in the signal processing community since it gives the best performance in term of Mean Square Error (MSE) at high Signal to Noise Ratio (SNR). Let $\hat{\omega}_m$ be an unbiased estimate of $\omega_m$, then
\begin{eqnarray}
E\{(\omega_m-\hat{\omega}_m)^2\} \geq [{\bf C}]_{mm} \stackrel{\rm def}{=} [ {\bf J}^{-1} ]_{mm}
\end{eqnarray}
where $E\{.\}$ is the mathematical expectation and ${\bf J}$ is the Fisher Information Matrix (FIM) defined by the Slepian-Bangs formula for a complex circular Gaussian observation ${\bf y} \sim \mathcal{CN}({\bf \bar{y}}, \sigma^2 {\bf I})$:
 \begin{eqnarray}
[{\bf J}]_{ij} &=&\frac{2}{\sigma^2} \Re\left\{ \left(\frac{\partial {\bf \bar{y}}}{\partial \omega_i}\right)^H \frac{\partial  {\bf \bar{y}}}{\partial \omega_j } \right\} = \frac{2}{\sigma^2} \sum_{t=1}^T \Re\left\{{\bf s}(t)^H \left(\frac{\partial {\bf A}}{\partial \omega_i}\right)^H \frac{\partial {\bf A}}{\partial \omega_j } {\bf s}(t)\right\}
\end{eqnarray}
where $ \frac{\partial {\bf A}}{\partial \omega_1} =
\begin{bmatrix} {\bf \dot{a}}(\omega_1) & {\bf 0} \end{bmatrix}$,
$ \frac{\partial {\bf A}}{\partial \omega_2} =  \begin{bmatrix}
{\bf 0} & {\bf \dot{b}}(\omega_2,\phi)  \end{bmatrix}$, $
\frac{\partial {\bf A}}{\partial \phi} =  \begin{bmatrix}  {\bf 0}
& {\bf \ddot{b}}(\omega_2,\phi)  \end{bmatrix}$ where $ {\bf
\dot{a}}(\omega_1) $ ({\em resp.} $ {\bf \dot{b}}(\omega_2,\phi)
$) is the first-order derivative of $ {\bf a}(\omega_1) $ ({\em
resp.} $ {\bf b}(\omega_2,\phi)  $) {\em w.r.t.} the parameters
$\omega_1$, $\omega_2$, respectively and $ {\bf
\ddot{b}}(\omega_2,\phi)$ is the first-order derivative {\em
w.r.t.} the parameter $\phi$. After some algebra, one obtains the
CRB for the parameters of interest and the coupling term:
\begin{eqnarray}\label{CRB1}
\mbox{CRB}(\omega_1) &=&  \frac{\sigma^2}{2} \frac{\beta }{ \mathcal{Q}}, \\
\mbox{CRB}(\omega_2)&=& \frac{\sigma^2}{2} \frac{ L_2 ||{\bf s}_1
||^2-\frac{\eta^2(\delta)}{L_3 || {\bf s}_2||^2}  }{
\mathcal{Q}},\\ \label{CRB3} \mbox{CRB}(\omega_1,\omega_2)&=& -
\frac{\sigma^2}{2} \frac{\zeta(\delta) - \frac{\eta(\delta)
L_3}{L_4} }{\mathcal{Q}}.
\end{eqnarray}
where  $\beta= ||{\bf s}_2 ||^2 \left( L_2 - \frac{L_3^2}{L_4}
\right)$, $\mathcal{Q} = \beta\left(L_2 ||{\bf s}_1
||^2-\frac{\eta^2(\delta)}{L_3 || {\bf
s}_2||^2}\right)-\left(\zeta(\delta)-\eta(\delta)
\frac{L_2}{L_3}\right)^2$, $L_r = \sum_{\ell=0}^{L-1} \ell^r $ and
$ \eta(\delta) =  \Re\left\{h  \sum_{\ell=0}^{L-1} \ell^2 e^{i
(\delta \ell + \phi \ell^2) }   \right\}$ and $\eta(\delta) =
\Re\left\{h \sum_{\ell=0}^{L-1} \ell^3 e^{i (\delta \ell +  \phi
\ell^2) } \right\}$ with $h = {\bf s}_1^H {\bf s}_2$.

\subsection{Derivation of the analytic ARL}

\subsubsection{The Smith's equation: }
The Smith's methodology \cite{smith} provides the ARL as the solution of the following equation:
\begin{eqnarray}\label{sm}
\mbox{CRB}(\delta) = \delta^2.
\end{eqnarray}

Expanding $\mbox{CRB}(\delta)$ {\em w.r.t.} the localization
parameters and the coupling term, we have \cite{nab}:
\begin{eqnarray}\label{sm0}
\mbox{CRB}(\delta) =  \mbox{CRB}(\omega_1) +\mbox{CRB}(\omega_2) -2 \mbox{CRB}(\omega_1,\omega_2)   \end{eqnarray}
where the CRBs are given by expressions (\ref{CRB1})-(\ref{CRB3}).

\subsubsection{Polynomial resolution of the linearized problem: }
Solving analytically equation (\ref{sm}) with (\ref{sm0}) and (\ref{CRB1})-(\ref{CRB3}) seems an intractable problem. Recalling that we assume that the separation $\delta$ is small then a first-order Taylor expansions of $\zeta(\delta) $ and $\eta(\delta) $ lead to the following approximations:
\begin{eqnarray}
\eta(\delta) \approx   \Re\left\{h  (v(\phi) +j \delta r(\phi))  \right\},\ \
\zeta(\delta) \approx   \Re\left\{h  (u(\phi) +j \delta v(\phi))  \right\}
\end{eqnarray}
in which $u(\phi)= \sum_{\ell =0}^{L-1} \ell^2 e^{i \phi \ell^2}$,
$v(\phi)= \sum_{\ell =0}^{L-1} \ell^3 e^{i \phi \ell^2}$ and
$r(\phi)= \sum_{\ell =0}^{L-1} \ell^4 e^{i \phi \ell^2}$. The
above relation can be rewritten in a linear form  {\em w.r.t.} the
separation according to
\begin{eqnarray}
 \Re\left\{h  (v(\phi) +j \delta r(\phi))  \right\} =  P - \delta Q,\ \
 \Re\left\{h  (u(\phi) +j \delta v(\phi))  \right\} =  P' - \delta Q'
\end{eqnarray}
where $P= \Re\{ h v(\phi)  \}$, $Q = -\Im\{h r(\phi)\}$, $P'=\Re\{
h u(\phi)  \}$ and $Q' = -\Im\{h v(\phi)\}$. Using these
approximations, one can rewrite the CRB as \begin{eqnarray} {\bf
C}  = {\bf J}^{-1} \stackrel{\rm def}{=}   \frac{\sigma^2}{2}
\begin{bmatrix} {\bf Q}^{-1} & \times \\ \times & \times
\end{bmatrix},
\end{eqnarray}
where, the  Schur complement can be approximated as follow $ {\bf
Q} \approx  \begin{bmatrix} P_2(\delta) & P_1(\delta)  \\
P_1(\delta) &  \beta \end{bmatrix}$ in which we have introduced
the two following polynomials $ P_2(\delta) = \alpha_1 \delta +
\alpha_0$ and $ P_2(\delta) = a_2 \delta^2 +a_1 \delta + a_0$
where $a_2 = -\frac{Q^2}{L_4 ||{\bf s}^2 ||^2}$,
$a_1=-\frac{2PQ}{L_4 ||{\bf s}^2 ||^2}$, $a_0=L_2|| {\bf s}_1
||^2-\frac{P^2}{L_4 ||{\bf s}^2 ||^2}$, $\alpha_1 =
Q'-\frac{L_3}{L_4} Q$ and $\alpha_0 = P'-\frac{L_3}{L_4} P$. The
linearized CRB expressions are now given by
\begin{eqnarray}\label{CRBl1}
\mbox{CRB}(\omega_1) Ê\approx   \frac{\sigma^2}{2} \frac{\beta }{
Q(\delta) },\\ \label{CRBl2}\mbox{CRB}(\omega_2)  \approx
\frac{\sigma^2}{2} \frac{ P_2(\delta) }{ Q(\delta)  },\\
\label{CRBl3} \mbox{CRB}(\omega_1,\omega_2)  \approx  -
\frac{\sigma^2}{2} \frac{P_1(\delta) }{ Q(\delta)},
\end{eqnarray}
where $
Q(\delta) = P_2(\alpha) \beta -P_1^2(\delta) = (\beta a_2-\alpha_1^2) \delta^2 +(\beta a_1 -2\alpha_0 \alpha_1) \delta + \beta a_0-\alpha_0^2$. Consequently from (\ref{sm}) and (\ref{CRBl1})-(\ref{CRBl3}), the Smith's equation becomes $
\delta^2 = \frac{Q'(\delta)}{Q(\delta)}$
where $
Q'(\delta)=\frac{\sigma^2}{2}( \beta+P_2(\delta) +2 P_1(\delta)) \stackrel{\rm def}{=}  c_2 \delta^2 + c_1 \delta +c_0$
with $c_2 = \frac{\sigma^2}{2} a_2$, $c_1 = \frac{\sigma^2}{2} (a_1+2\alpha_1)$ and $c_0= \frac{\sigma^2}{2}(\beta+a_0+2\alpha_0)$. So, it is easy to see that the Smith's equation provides the ARL as the solution of a 4-th order polynomial according to
\begin{eqnarray}\label{poly1}
R(x) = Q(x) x^2 - Q'(x)  \stackrel{\rm def}{=}  x^4 + g_3 x^3+g_2 x^2 + g_1 x +g_0
\end{eqnarray}
where $x$ is a free variable with $
g_0 = -\frac{c_0}{\beta a_2 - \alpha_1^2},$ $
g_1 = -\frac{c_1}{\beta a_2 - \alpha_1^2}$, $
g_2 = \frac{\beta a_0 -\alpha_0^2-c_2}{\beta a_2 - \alpha_1^2},$ $
g_3 =  \frac{\beta a_1 -2 \alpha_0\alpha_1}{\beta a_2 - \alpha_1^2} $
where we assume $\beta a_2 \neq \alpha_1^2$.


\subsubsection{Analytic solutions of $R(x)$:}

The resolution of a 4-th order polynomial is not straightforward
but we propose a solution to this problem. More precisely, as
$\mbox{CRB}(\delta) = \mbox{CRB}(-\delta)$, it is easy to see that
if $\delta$ is a solution of the Smith's equation then $-\delta$
is also a solution. This implies that if $\delta$ is a root of
$R(x)$ then $-\delta$ is also a root. So, $R(x)$ has four roots,
namely $\{\delta, -\delta, r_1, r_2\}$ and a decomposition of this
polynomial into a product of monomial terms is given by
\begin{eqnarray}
R(x) &=&(x-\delta)(x+\delta)(x-r_1)(x-r_2) \\ &= & x^4
-(r_1+r_2)x^3 + (r_1r_2-\delta^2) x^2\\ \label{poly2} &+&
\delta^2(r_1+r_2) x - r_1r_2\delta^2.
\end{eqnarray}

We can identify the coefficients of the polynomials (\ref{poly1})
and (\ref{poly2}) according to
\begin{eqnarray}\left\{
\begin{array}{ccc}
-(r_1+r_2) &=&  \frac{\beta a_1 -2 \alpha_0\alpha_1}{\beta a_2 - \alpha_1^2},\\
r_1r_2-\delta^2 &=& \frac{\beta a_0 -\alpha_0^2-c_2}{\beta a_2 - \alpha_1^2},\\
\delta^2(r_1+r_2) &=& -\frac{c_1}{\beta a_2 - \alpha_1^2},\\
-r_1r_2\delta^2 &=&  -\frac{c_0}{\beta a_2 - \alpha_1^2}.
\end{array}\right.
\end{eqnarray}

Combining the second and the last equations, we have to solve the
following polynomial:
\begin{eqnarray}
R'(x) = (\beta a_2-\alpha_1^2) x^4+(\beta a_0-\alpha_0^2-c_2)x^2
-c0.
\end{eqnarray}

Polynomial $R'(x)$ can be reformulated as a 2-rd order polynomial
according to
\begin{eqnarray}
R'(z) = (\beta a_2-\alpha_1^2) z^2+(\beta a_0-\alpha_0^2-c_2)z
-c0.
\end{eqnarray}

The discriminant is
\begin{eqnarray} \label{Dis}
\Delta  = (\beta a_0-\alpha_0^2-c_2)^2\left( 1+4 \frac{(\beta
a_2-\alpha_1^2) c_0}{(\beta a_0-\alpha_0^2-c_2)^2}\right).
\end{eqnarray}

The study of the sign of the discriminant is not straightforward
but observe that $c_0, c_2 \stackrel{\sigma^2 \rightarrow
0}{\longrightarrow}  0$, so if the noise variance is low then the
discriminant is given by
\begin{eqnarray}\label {discri}
\Delta  \stackrel{\sigma^2 \rightarrow 0}{\longrightarrow} (\beta
a_0-\alpha_0^2-c_2)^2 \geq 0.
\end{eqnarray}

So, we know that if the noise variance is not too high, there
exists two candidates for the ARL (or a double solution if the
discriminant is zero) which are given by
\begin{eqnarray} \label{ARLpossible}
\delta = \sqrt{\frac{-(\beta a_0-\alpha_0^2-c_2)\pm \sqrt{\Delta}
}{2(\beta a_2-\alpha_1^2)}}\end{eqnarray} where $\sqrt{\Delta}$ is
given by the square root of expression (\ref{Dis}).

\subsubsection{Analytic expression of the ARL:}

To discriminate the two possible solutions, we advocate that
$\delta \stackrel{\sigma^2 \rightarrow 0}{\longrightarrow}  0$  is
a reasonable property. Thus, note thanks to (\ref{discri}),  the
sign in (\ref {ARLpossible}) must be chosen as positive to ensure
$\delta \stackrel{\sigma^2 \rightarrow 0}{\longrightarrow}  0$. If
not (in case of negative sign in (\ref {ARLpossible}))  the chosen
solution  will be meaningless. So, finally the ARL is given by
\begin{eqnarray} \label{ARL}
\delta = \sqrt{\frac{-(\beta a_0-\alpha_0^2-c_2)+\sqrt{\Delta}
}{2(\beta a_2-\alpha_1^2)}}.\end{eqnarray}

To provide further simplifications, we can see that
\begin{eqnarray}
\sqrt{\Delta}  \approx  \beta a_0-\alpha_0^2-c_2 +  \frac{2(\beta
a_2-\alpha_1^2) c_0}{\beta a_0-\alpha_0^2-c_2}
\end{eqnarray}
thanks to a first-order Taylor expansion of the square root
$\sqrt{1+x} \approx 1+1/2 x$ for small $x$. Using the above
approximation of $\sqrt{\Delta}$, the ARL takes the simple
following expression:
\begin{eqnarray} \label{ARLe}
\delta &\approx& \sqrt{\frac{c_0}{  \beta a_0-\alpha_0^2-c_2 }}.
\end{eqnarray}

So, we  can see that  $\delta \approx O(\sigma)$.

\section{Numerical illustrations}

\begin{figure}[htb]
\begin{center}
\begin{tabular}{cc}
\includegraphics[width=8cm,height=6cm]{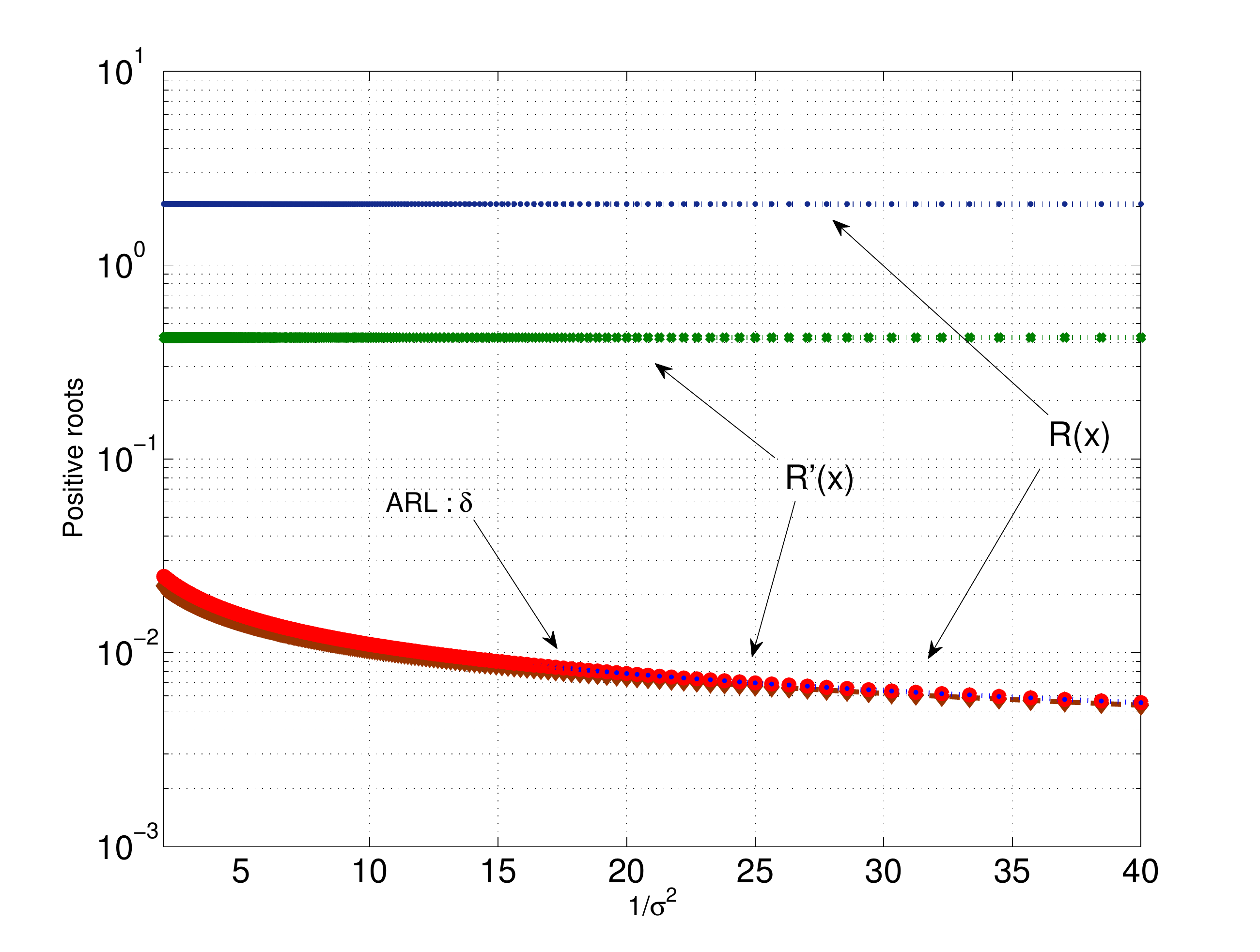}&
\includegraphics[width=8cm,height=6cm]{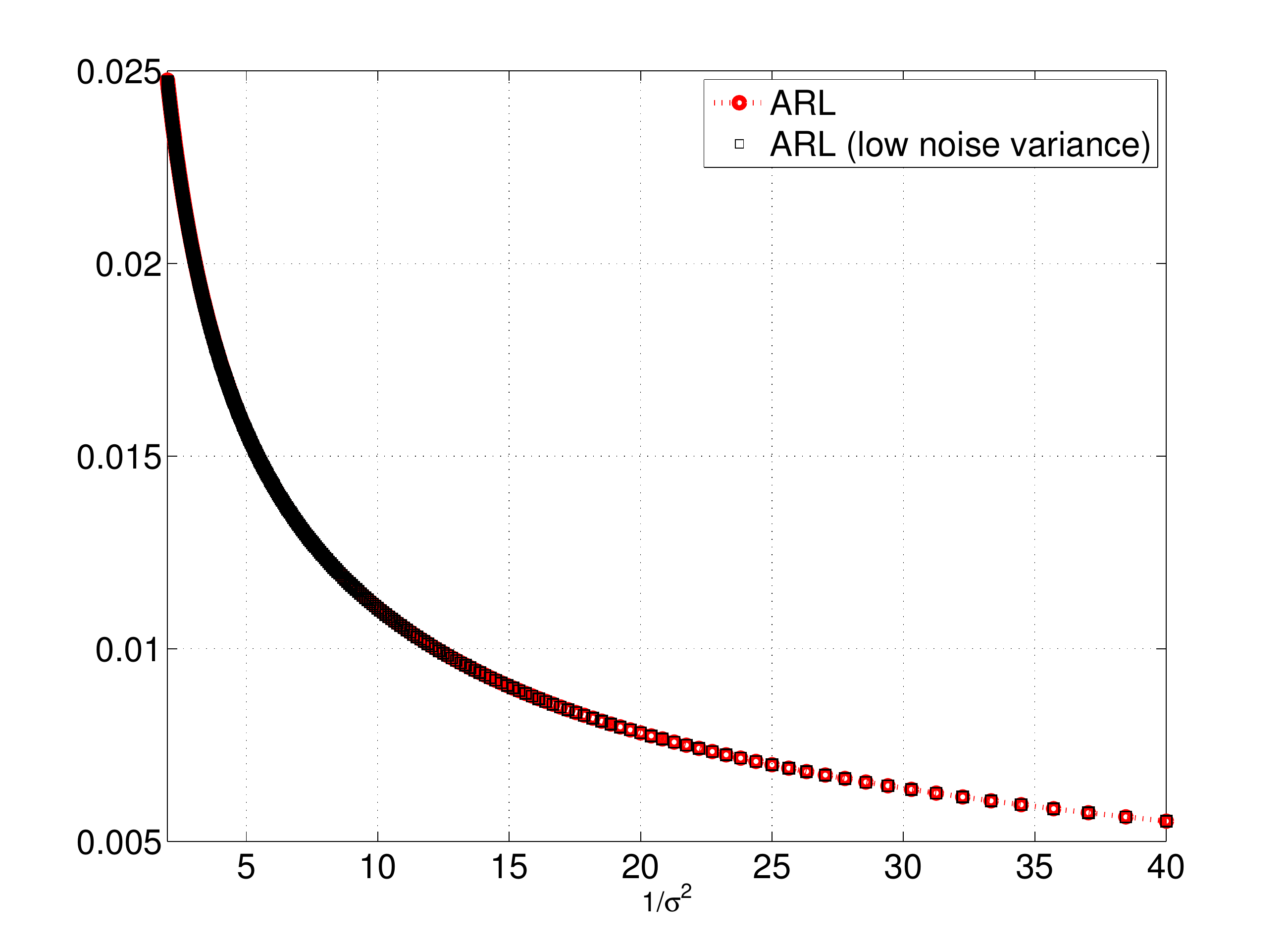}
\end{tabular}
\caption{(a) Positive roots for polynomials $R(x)$, $R'(x)$ and analytic ARL given in expression (\ref{ARL}) vs. the inverse of the noise variance. (b) ARL given in  expression (\ref{ARL}) and the approximated ARL under the assumption of low noise variance  given in expression (\ref{ARLe}) vs. the inverse of the noise variance.}\label{fig1}
\end{center}
\end{figure}

In this simulation part, we have considered an array constituted
by $L=10$ sensors. The sources are chosen to be close where
$\theta_{FF}=\pi/3$ and $\theta_{NF}=\pi/3.1$ and the range of
this source belongs to the interval $[0.62 (d^3
(L-1)^3/T)\lambda,2 d^2 (L-1)^2/\lambda     ]$ \cite{abed} where
$d=0.0125$ m and the carrier frequency is $f_0=10$ Mhz (the
wavelength is $\lambda = c/f_0$ where $c$ is celerity of the
light). The modulus of the NF sources is fixed to  ten times
higher than the modulus of FF source. It is normal to assume that
the source which is the closest has a higher power than the one
which belongs to the FF. The number of snapshot is $T=100$. On
Fig. \ref{fig1}, we have reported the positive roots of each
polynomials $R(x)$, $R'(x)$ and the analytic ARL given in
expression (\ref{ARL}). As we can see one root for polynomials
$R(x)$ and $R'(x)$ are independent from the noise variance. These
two roots have to be ignored. In addition, we can see that the one
root for $R(x)$ and $R'(x)$ are equal and follow a decreasing
function {\em w.r.t.} the inverse of the noise variance. Moreover,
the analytic ARL given in expression (\ref{ARL}) is in good
agreement with these roots and thus assess the validity of  the
derivations given in the previous section. The ARL given in
expression (\ref{ARL}) and the approximated ARL under the
assumption of low noise variance  given in expression (\ref{ARLe})
are reported on Fig. 2. It is important to highlight the good
accuracy of the proposed approximated ARL.

\section{Conclusion}

In this paper, we have derived and analyzed the Angular Resolution Limit (ARL) based on the resolution of the linearized Smith's equation for the new and realistic scenario where far-field and near-field sources are mixed. Generally speaking, the ARL is important and fundamental since this quantity gives the limit in the resolvability/separation of two closely spaced signals in term of their direction of arrivals. We show that for the chosen application, the resolution of the Smith's equation turns to be  the selection of the "right" root of a 4-th order polynomial.   This allows us to give a closed-form (analytic) expression of the ARL.

\end{document}